\documentclass[%12pt,
amsmath, aps, prl, twocolumn ]{revtex4}
\usepackage{bm}
\usepackage{graphicx}
\usepackage{amssymb}
\def \d{\partial}
\def \bv{{\bf v}}
\def \bu{{\bf u}}
\def \br{{\bf r}}
\def \bw{{\bf w}}
\def \brho{\boldsymbol{\rho}}

\def \bomega{\boldsymbol{\omega}}

\def \tr{\mbox{tr}}

\begin{document}
\title{On the multifractal structure of fully developed turbulence.}
\author{K.P.Zybin\footnote{Electronic addresses: zybin@lpi.ru, sirota@lpi.ru}
and V.A.Sirota$^*$}
\affiliation %\institute
{P.N.Lebedev Physical Institute of RAS, 119991, Leninskij pr.53,
Moscow, Russia }
%\title{Vortex filaments and multifractal structure of fully developed
%hydrodynamical turbulence.}
\begin{abstract}
The appearance of vortex filaments, the power-law dependence of
velocity and vorticity correlators and their multiscaling behavior
are derived  from the Navier-Stokes equation. This is possible due
to interpretation of the Navier-Stokes equation as an equation with
 multiplicative noise, and remarkable properties of random matrix products.
\end{abstract}
\maketitle

\section{Introduction}

Understanding of statistical properties of a turbulent flow is a classical problem of hydrodynamics. Intermittent behavior of velocity scaling exponents demonstrated in experiments and numerical simulations \cite{observations,simulations, arneodo} is one of its important aspects.  The most successful, and conventional nowadays, way to interpret this intermittency is the Multifractal (MF) approach introduced in \cite{ParisiFrisch}. This is  a generalization of Kolmogorov's K41 theory, and it allows to express   all the observed intermittent characteristics by means of one function $D(h)$. In particular, if $D(h)$ is derived from observed velocity scaling exponents, then all the other values can be calculated \cite{25years}.

However, the phenomenological presentation of the MF model assumes the existence of singularities (actually, the paper \cite{ParisiFrisch} was titled 'On the singularity structure of fully developed turbulence'). The question if singularities can appear after a finite time has been widely discussed and remains still open \cite{Frisch, Kuznetsov, Kuznetsov2}. But in the absence of even one example of a finite-time singularity, the existence of a whole 'spectrum' of singularities is rather doubtful.

To make the MF theory independent of this assumption,  its probabilistic reformulation was introduced \cite{Frisch, MeneveauSreeni91, Mandelbrot91}. However, this formulation does not allow to  understand what happens in a turbulent flow; the structures (or the solutions to the Navier-Stokes equation) that are responsible for the observed intermittent properties, remain unknown.

On the other hand, many experiments and numerical simulations have shown the presence of 'coherent structures' - vortex filaments and 'pancakes' - in a turbulent flow (see, e.g., \cite{Frisch208vnizu or others?}). In \cite{Farge} it was shown that these structures make a fundamental contribution to the observed Kolmogorov's two-thirds law, and they contain the most part of the whole enstrophy of a flow. The origin of these coherent structures is still obscure (in \cite{Frisch} the tendency of a flow to produce these structures was characterized as 'mysterious'). Understanding of these filaments formation seems to be very important as it may also help to understand the cause of intermittency and multifractality.

In our previous papers we have developed a model of vortex filaments (VF). %\cite{JETP1,PRL1,PRE}.
We showed that it did not contradict to the MF model
\cite{PhysicaD}. In \cite{PRE2,notPRE,Scripta13} we combined the VF
theory with the MF approach; this allowed to calculate both
longitudinal and transverse velocity scaling exponents without
adjusting parameters, the result agreed well with numerical data.
Although this model was based on the Navier-Stokes equation (NSE),
it still contained an additional supposition: it assumed that the
nonlinear part of the pressure hessian was orthogonal to the local
vorticity inside the vortex filaments. This assumption was confirmed
by reasonable considerations but it was not proved thoroughly.

In this paper we propose a new accurate approach that may help to derive the multifractality
and the existence of vortex filaments based on the NSE. In our approach, the power spectrum
appears as a result of stretching of a vortex filament. Scaling exponents of different orders
 are produced by different filaments contributing
mostly to these orders. No finite-time singularities are needed,
singularity is reached (in non-viscous limit) after infinite time.
The power-law dependence of velocity in the vicinity of  the
forming singularity (not the cascade of decaying  vortices)
produces the long power-law tail of the Fourier spectrum.

A multiplicative noise instead of additive random forcing was used
in papers concerned to dynamics of passive scalar and vector fields
\cite{Zeldovich,Falkovich,Antonov}. We use this approach
 to introduce randomness into the NSE or Euler equation; this allows us to consider
the evolution of  small-scale velocity perturbations in the 'external' field produced
by random large-scale velocity (the next Section). We find a long-time asymptotics of
general solution for small-scale fluctuations (Section 'Asymptotic analysis \dots ').
It appears that this asymptotics depends on some combination of large-scale values,
which is random, but tends to a constant as time increases. Analyzing the solution,
we observe an effect similar to that described in \cite{Frisch} on the basis of numerical
 simulations: in some special reference frame, velocities depend  to leading order on only
 one coordinate, thus the flow becomes one-dimensional; this causes a depletion of
 nonlinearity. The evolution along this one coordinate corresponds to stretching
 of the rotating vortex filament.

In the next Section  we thus introduce a simplified model, in which the large-scale field is fixed and  the special reference frame coincides with the laboratory frame. This particular case allows to understand better the details of solution, to see what happens to the spectrum,  and to estimate the effect of  viscosity.

In the Section 'Introduction of stochastics'  we discuss the first-order correction to the solution; not only averages of the combined large-scale value mentioned above, but also fluctuations around them are now  taken into account. We see that fluctuations of the large-scale flow result in multifractality of small-scale correlators.

In the last Section we discuss the results of the paper.

\section{  Equation for small-scale velocity fluctuations}

A turbulent flow can, in principle, be completely described
 by the dynamical NSE. The pulsations appear as a result of
 instability of the flow. But solving the complete problem with
 account of instabilities seems to be impossible. So,
 to derive statistical properties of the flow, one has to introduce
randomness into the  equation. This is usually done by adding a
large-scale random external force  into the right-hand side of the
NSE:
\begin{eqnarray}\label{0}
\frac{\partial \bv}{\partial t} + (\bv \nabla) \bv &=& -\nabla P +
{\bf F}({\bf r},t) + \nu \Delta {\bf v} \,, \\ \nonumber
 \nabla \cdot {\bf v} &=& 0
\end{eqnarray}
The  probability distribution of the force is usually supposed to
be Gaussian. One assumes that the resulting correlation properties
inside the inertial range do not depend on the properties of the
exciting force. To satisfy this requirement, $\bf F$ must not
include small-scale pulsations: not only its correlator must decay
at scales of the order of largest eddies' turnover scales $L$, but
also the realization of $\bf F$ must consist of large-scale
harmonics only.

But the external volume-acting forces might exist in the flow and
might not. On the other hand, this way does not allow to separate
large-scale and small-scale velocity fluctuations; introducing the
large-scale force does not make any simplification. %y the equation.

An alternative way is to introduce stochasticity as a multiplicative
noise; it is used in papers on passive scalar and vector fields
(e.g., \cite{Zeldovich,Antonov,Falkovich}). We apply this approach
to the NSE and consider
%We propose a new approach to the problem:
large-scale velocity perturbations (not forces) % are considered
as given random process. The stochastic properties of small-scale
fluctuations can then be derived based on the properties of the
large-scale fluctuations.

\subsection{ Formal introduction of randomness}

To develop this idea, introduce a random  field $\tilde{ \bf
U}$. To make it large-scale in the sense discussed above,
 we smoothen  $\tilde{U}$  by means of a space average
over surrounding volume of the order of $L$, e.g.:
\begin{equation}    \label{smoothing}
{\bf U}({\bf r},t) = \frac 1{L^3} \int \tilde{\bf U}(\br+\brho,t)
e^{-\rho^2/L^2} d\brho
\end{equation}
and we require  $\nabla \cdot {\bf U} =0$.

Now we {\it define} the large-scale force $\bf F$ and pressure $\pi$ according to:
\begin{eqnarray} \label{LargeNew}
 \frac{\partial {\bf U}}{\partial t} + ({\bf U} \nabla) {\bf U} &=&
-\nabla \pi + {\bf F}({\bf r},t) + \nu \Delta {\bf U} \,,\\
 \nabla \cdot {\bf F} &=& 0 \nonumber
\end{eqnarray}
It is evident that $\bf F$ is large-scale and satisfies the above
condition. (To avoid possible divergences of the time derivative, one
often  considers generalized  solutions; as we will
see later on, this does not make any difference since only   $\bf U$ and its integrals, not derivatives,
contribute to the result.)

We then substitute this $\bf F$ to the right-hand side of (\ref{0}),
and seek the solution in the form
$$
 \bv ({\bf r},t) = {\bf U} + {\bf u} \ , \qquad P=p+\pi
$$
Then for $\bf u$ we get the equation:
\begin{eqnarray}
%\begin{array}{l} \displaystyle
& \frac{\partial}{\partial t} u_i + ({\bf U}\nabla) u_i + ({\bf u}
 \nabla) U_i +({\bf u}\nabla )u_i
 = - \nabla_i p   + \nu \Delta u_i \,,& %\quad
 \nonumber
 \\ & \nabla {\bf u} = 0 &  \label{maineqwithU}
%\end{array}
\end{eqnarray}
We regard this equation as the stochastic version of the NSE with stochasticity
introduced by means of the large-scale random velocity field $\bf U$
(instead of random force).

We note that $\bf F$ is not presented in this resulting equation.
So, in this approach it is not important if the exciting force does
exist or not. In this sense, introduction of probability by means of the
large-scale velocity field $\bf U$  includes the
NSE with external forces as a particular case.

As it was with large-scale forces, we assume that any particular
choice of $\bf U$ does not affect significantly the statistics
 of the flow at small scales. Indeed, as we will see
below, the choice of the statistical properties of $\bf U$
%, and in particular its gaussianity,
is not important for the existence of scaling exponents.

%We stress that no separation of  scales is needed; ${\bf u}$ may include all scales,
%although in what follows we will be interested in its small-scale component.

\subsection{    The small-scale limit}

We now simplify (\ref{maineqwithU}) to analyze it analitically. To this
purpose, we  consider a small vicinity of some point.
 The smoothed function $\bf U$ can be expanded in a Taylor series for
$r\ll L$, so
\begin{equation} \label{seriesV}
U_i({\bf r},t) = U_i(0,t) + A_{ij}(t) r_j + \dots \ , \quad \mbox{tr} A =0
\end{equation}
The omitted terms are smaller by a factor $\sim r/L$.
 We hereafter restrict our consideration to the first two terms since  in what follows
 we will be interested in small scales.  This corresponds to taking the limit $L
\to \infty$ with large-scale eddy turnover time $T$ remaining
constant. Note also that the drift component ${\bf U}(0,t)$ can
easily be taken zero by  choosing the appropriate reference frame.

Then from (\ref{maineqwithU}) we obtain:
\begin{eqnarray}
 \frac{\partial}{\partial t} u_i + (A_{kj}r_j\nabla_k) u_i &+&
A_{ik}u_k +({\bf u}\nabla )u_i  \nonumber \\
&=& - \nabla_i p   + \nu \Delta u_i \,,  \label{maineq} \\
  \nabla_i u_i &=& 0  \nonumber
\end{eqnarray}
This is the main equation of the paper.  In the limit $r\ll L$, it
is an exact consequence of the Navier-Stokes equation. The
large-scale velocity gradients $A_{ij}$ play the role of external
forces, and, instead of the forces, they are used to introduce
stochasticity into the equation. Based on their statistical
properties, one can now analyze the equation to find statistical
properties at small scales.

To restrict our consideration by scales $l\ll L$, let now the initial velocity
field $\bf u$ have scales at least several times smaller than  $L$.Then (\ref{maineq})
would be valid if the characteristic scale of $\bf u$ became smaller during the evolution,
and would fail if the scale of $\bf u$ increased. As we will see below, the first possibility
takes place in our solution: the distribution of $\bf u$ becomes narrower and sharper, producing peaks.
So the validity of the approximation $l\ll L$ improves exponentially.

Our task is now to study the asymptotic properties of the stochastic
equation (\ref{maineq}) after long time.

\section{Asymptotic analysis of (\ref{maineq}): inviscid limit}

We now consider the Euler analog to (\ref{maineq}), i.e. $\nu =0$
(the contribution of viscosity will be discussed later).

First, in order to eliminate the two linear terms we change the variables $\br,\bu$ to ${\bf X}, \bw$ according
to
\begin{equation} \label{zamena}
u_i(r, t) = g_{i\mu}(t) w_{\mu}(X_{\nu}, t)\,,\quad X_{\nu} =
q_{\nu j}(t) r_{j}
\end{equation}
where $g_{i\mu}(t)$ and $q_{\nu j}(t)$ satisfy the equations:
\begin{eqnarray}  % \label{g}
\dot{g}_{i\alpha} + A_{ij} g_{j\alpha} = 0\,,\qquad
g_{i\alpha}(0)=\delta_{i \alpha}
\nonumber \\ %  \end{equation}   \begin{equation}
\label{matricy} %\\ \nonumber
% \label{q}
\dot{q}_{\gamma j} + q_{\gamma k}
A_{k j} =0 \,, \qquad q_{\alpha j}(0) =\delta_{\alpha j}
\end{eqnarray}
Substituting to (\ref{maineq}) we get:
$$
g_{i\mu}\left( \frac{\partial w_{\mu}}{\partial t} +
q_{\lambda j} g_{j\alpha} w_{\alpha} \frac{\partial
w_{\mu}}{\partial X_{\lambda}}\right) = - q_{\nu i} \frac{\partial
p}{\partial X_{\nu}} \ ,
$$
$$
\quad q_{\nu i} g_{i\mu}
\frac{\partial w_{\mu}}{\partial X_{\nu}} = 0
$$
In this paper  we restrict ourselves by the consideration of
symmetric $A_{ij}$: we will discuss the inner parts of vortex
filaments where vorticities are very high, so we expect that small
'external' large-scale vorticity (which is equal to the asymmetric
part of $A_{ij}$) does not play a crucial role.
From  $A=A^T$ %then from (\ref{g}) and (\ref{q}),
it follows
$$g_{j \alpha}=q_{ \alpha  j}$$
The equation then becomes
\begin{eqnarray} \nonumber
&& \frac{\partial w_{\mu}}{\partial t} + q_{\lambda j}
g_{j \alpha} w_{\alpha} \frac{\partial w_{\mu}}{\partial
X_{\lambda}} = -  \frac{\partial p}{\partial X_{\mu}} \ , \\
%\qquad
\label{withqg} && \\
&& \quad q_{\nu i} g_{i\mu}  \frac{\partial w_{\mu}}{\partial
X_{\nu}} = 0 \nonumber
\end{eqnarray}
 Since $A_{ij}$ is a
random process, the matrices $g_{i \alpha}$ and $q_{\alpha j}$ are also random.
Note that only the combination $q_{\nu i}g_{i \mu}$ is presented in
(\ref{withqg}). To analyze the solutions of (\ref{withqg}) at $t\to
\infty$, we need to know the asymptotic behavior of this value.

\subsection{Asymptotic behavior of $q,g$ and asymptotic solution of
(\ref{withqg})}

 To examine the
solution of (\ref{matricy}), we proceed to a  discrete
approximation: consider a discrete sequence of moments separated by
$\Delta t$ and let $A_{ij}(t)=(A_n)_{ij}$ be constant  inside each
small ($n$-th) interval. Then, for each $\Delta t$, the solution to
the Eq. (\ref{matricy}) is described by an exponent, and we get
$$ q_n = q_{n-1} e^{-A_{n} \Delta t} $$
Hence,
\begin{equation} \label{qsequence}
 q_N = e^{-A_1 \Delta t}\cdot e^{-A_2 \Delta t}\cdot \dots \cdot
 e^{-A_N \Delta t}
\end{equation}
We now consider the Iwasawa decomposition of the matrix $q_N$:
\begin{equation} \label{Iwasawa}
 q = z(q) d(q) s(q)
\end{equation}
 where $z$ is an upper triangular matrix with diagonal elements
equal to 1, $d$ is a diagonal matrix with positive eigenvalues, $s$
is an orthogonal matrix.

 The matrix $q_N$ is a multiplication of $N$ random real unimodular matrices with the
same distribution. The asymptotic behavior of this object has been
studied carefully, and a number of important results has been
obtained. (For short summation of them, see \cite{Letchikov}.)  In
particular, the following Theorems have been proved under reasonable
conditions: %
$^2$\footnotetext[2]{Note that the Theorems assume neither
symmetry nor gaussianity of $A$.}%
\begin{enumerate}
\item  \cite{Let18}  with probability 1, there exists the limit $\lim
\limits_{N \to \infty} \frac 1N \ln d_i (q_N) = \lambda_i$,
$\lambda_i$ are  not random, i.e., do not depend on the realization of the process $A_{ij}(t)$
 but only on the statistical properties of the process;
$\lambda_1<\lambda_2<\lambda_3$, the ordering is due to the triangular
matrix which provides the inequality   of the axes;

\item \cite{Let9, Let13} the distribution of $\xi_i = \frac
{ \left( \ln d_i (q_N) - \lambda_i N \right)}{\sqrt{N}}$ is
asymptotically close to a Gaussian distribution and (weakly)
converges to it as $N\to\infty$;

\item \cite{Let11} with probability 1, $z(q_N)$ converges as $N\to
\infty$, $z(q_N) \to z_{\infty}$; oppositely to $\lambda_i$,  the values $z_{\infty}$ are different
in different realizations of $A_{ij}(t)$.

\item \cite{Let10} the values $\xi _i (q_N)$ and  $z(q_N)$ are
asymptotically independent.

\end{enumerate}

For our purposes, these results can be written shortly as
 $^3$\footnotetext[3]{The transition to the limit $N\to\infty$
 corresponds to $t\to \infty$ with $\delta t $ remaining constant.
 To return back from discrete to continuous description, one has to
 take the limit $\Delta t \to 0$ afterwards. This would result in
 replacement $N$ by $t$ and renormalization of $\lambda_i$. The exponents in
 (\ref{zdasymptote}) would take the form $\lambda_i t + \int \xi_i dt$.   }
\begin{eqnarray}
&& z(q_N) \to z_{\infty} \ , \nonumber \\
 \nonumber
 && d (q_N) = \mbox{diag} %\left
 (
e^{\lambda_1 N+ \sum \limits_n^N \xi_1(n)},e^{\lambda_2 N+ \sum
\limits_n^N \xi_2(n)}, e^{\lambda_3 N+ \sum \limits_n^N \xi_3(n)}
%\right
)  , \\ %\quad
&&\lambda_1<\lambda_2 <\lambda_3  \label{zdasymptote}
\end{eqnarray}
  Note that % the matrix $z_{\infty}$  and
the coefficients $\lambda_i$ are constants determined by statistical properties of
the random process $(A_n)_{ij}$; $z(q_N)$ have the limit $z_{\infty}$ that depends on the realization.
 To the contrary, $s(q_N)$ changes
quickly as a function of $N$, and depends strongly on the
realization of $A_n$.

In the case $A=A^T$, fortunately, the rotating term vanishes in the
combination $qg$:
$$
(q g)_N  = (q q^T)_N \simeq  z_{\infty} d^2 (q_N) z^T_{\infty} \ ,
$$
$$ d^2 (q_N)= e^{2\lambda_3 N} \cdot \mbox{diag}(0,0,1) + O(e^{2\lambda_2
N})
$$
Neglecting the terms growing slower than $e^{2\lambda_3 N} $, we get
$$
(q g) _N  = C e^{2\lambda_3 N}
$$
where, with accuracy $O \left( e^{2(\lambda_2-\lambda_3)N} \right)$,
$C$ is a constant symmetric matrix $C=z_{\infty} diag(0,0,1)
z^T_{\infty}$.
%Rotating the coordinate frame, one can make it diagonal.

We now  introduce a new vector variable ${\bf W} = C {\bw}$
instead of $\bw$, and we return to the continuous description
$qg(t)$. Then from (\ref{withqg}) we get
$$
\frac{\partial {\bf W}}{\partial t} + e^{2\lambda_3t}\left({\bf
W}\frac{\partial}{\partial {\bf X}}\right) {\bf W} = - C
\frac{\partial p}{\partial {\bf X}}\,,\qquad
 \frac{\partial{\bf
W}}{\partial {\bf X}} = 0
$$
From  (\ref{seriesV}), (\ref{qsequence}) it follows that $\det q_N =1,$ and hence    $\lambda_1+\lambda_2+\lambda_3 = 0$; thus,  $\lambda_3 > 0$. Thus, the asymptotic ($t\to \infty$)
solution to the first order by $e^{-2\lambda_3 t}$ takes the form:
\begin{equation}\label{reshenie}
\left({\bf W}\frac{\partial}{\partial {\bf X}}\right) {\bf W} = - C
\frac{\partial}{\partial {\bf X}}\Pi \,,\quad \frac{\partial{\bf
W}}{\partial {\bf X}} = 0\,,\quad p=e^{2\lambda_3t} \Pi
\end{equation}
This equation is the asymptote of Eq.(\ref{withqg}) at large values
of $t$. We note that the constant matrix $C_{ij}$ is the only
remnant of the random process $A_{ij}(t)$  in this equation. This is
of course due to the chosen variables $({\bf X}, {\bf W})$; the
randomness remains in rotation of the corresponding reference frame.

The matrix $C$ is symmetric and can be reduced to diagonal by some
(time-independent) twist of the reference frame.  Thus, the
solutions of (\ref{reshenie}) correspond to some stationary
hydrodynamical configurations.

So, we see that the relations
(\ref{zamena}),(\ref{matricy}),(\ref{Iwasawa})-(\ref{reshenie}) set
up  a correspondence between general stochastic solutions of Eq.
(\ref{maineq}) (with symmetric $A_{ij}$) at very large values of $t$ ($t\to \infty$) and
stationary solutions of the Euler equation.

\subsection{Analysis of the solution}

We do not specify the solution of (\ref{reshenie}), assuming it to
be some general function $W(x)$. To understand the properties of the
solution (\ref{reshenie}), we have to rewrite it back in laboratory
coordinates $(\br,\bu)$. With account of
(\ref{zamena}),(\ref{matricy}),(\ref{Iwasawa}) we have
$$
\bu = g {\bf w} = % g C^{-1} {\bf W}(X) =
q^T C^{-1} {\bf W}({\bf X}) = s^T d \, z^T C^{-1}
{\bf W}({\bf X}) \ ,
$$
$$  {\bf X}=q \br = z d s \br
$$
To separate the stochastic rotational part of the solution, we make
one more change of variables:  %pass to one more reference frame
\begin{equation} \label{rotation}
 \br' = s \br \ , \qquad \bu' = s \bu
\end{equation}
This frame rotates randomly, since the matrix $s$ is a stochastic
function of time (as opposed to $z$ and $d$, which tend to a
constant or change steadily at large $t$). Also for any ${\bf
W}({\bf X})$  %satisfying
determined by (\ref{reshenie}), we define a new
vector function
$$
{\bf V}({\bf y})=z^T C^{-1} {\bf W}(z {\bf y})
$$
Then
$$
\bu' = d \, {\bf V} (d \, \br')
$$
or, in more detailed writing,
$$
u'_i = e^{\lambda_i t} V_i (e^{\lambda_1 t} r'_1, e^{\lambda_2 t}
r'_2 , e^{\lambda_3 t} r'_3)
$$
(no summation is assumed).

We see that in the rotating coordinates $\br'$, the asymptotic
solution is not random.

As $t\to \infty$, %the third argument of $V_i$ becomes much larger
%than the others, and
the third component $u_3'$ dominates, and the solution stretches
exponentially with different coefficients along different axes.
% This corresponds to exponential stretching along the third axis.%
%
Hence, to the leading order it is enough to account  the dependence
of $\bf u$  % OR:  to treat $\bf u$ as depending
on only one ($r'_3$) variable.%
$^4$\footnotetext[4]{ From the condition $\nabla\cdot \bu =0$ it
follows
$$\frac{\partial}{\partial \br'}\cdot \bu' = \sum \limits_i
 e^{2\lambda_i t}\frac{\partial V_i}{\partial y_i} (e^{\lambda_1 t} r'_1, e^{\lambda_2 t}
r'_2 , e^{\lambda_3 t} r'_3) =0 \ ,
$$
hence $\partial V_3 / \partial r_3 =0$.
}

We now take the curl to find  vorticity:
$$ \omega'_k = \varepsilon_{kji} \frac{\partial u'_i}{\partial r'_j } =
\varepsilon_{kji} e^{\lambda_i t} \frac{\partial V_i}{\partial y_j}
e^{\lambda_j t}
$$
Since  $\sum \lambda_i=0$, we have $\omega_k \propto e^{-\lambda_k
t}$. Hence, vorticity is directed mainly along the $r'_1$ axis:
\begin{equation} \label{solut}
\omega' \simeq \omega'_1 = e^{-\lambda_1 t} f \left(  e^{\lambda_3
t} r'_3 \right)
\end{equation}
We note that, since $\bomega '=s \bomega$, the absolute values of
vorticities are equal in the two frames, so $\omega=\omega'$.

Thus, vorticity (and velocity) is transported from boundaries to the
center, and simultaneously it grows exponentially. To keep the whole
system stable, we have to demand that at some point, e.g. $r_3'=L$,
vorticity (or velocity) is nearly constant:
\begin{equation} \label{granomega}
 \omega(t,L) \sim 1
 \end{equation}
In fact, the requirement is much weaker; it would be enough for
$\omega(t,L)$ not to grow exponentially. This is true for a point of
general position.%
 $^5$\footnotetext[5]{ Indeed, if the measure of the points where
vorticity grows exponentially with time is  $\mu > 0$, then velocity
also grows exponentially in the region. Then, to satisfy the
condition that energy does not grow exponentially on average, $\mu$
must decrease exponentially itself. }
 However, here we ask $\omega$ to be nearly constant at the boundary for simplicity. We shall
discuss the subject in the next Section.

With account of the boundary condition, we have $f(e^{\lambda_3
t'}L) \sim e^{\lambda_1 t'} $ for any $t'$. Choosing $t'$ in such a
way that $e^{\lambda_3 t} r'_3=e^{\lambda_3 t'}L$, we rewrite the
solution (\ref{solut})  in the form:
\begin{equation} \label{answer}
\omega(t,r'_3) \propto \left( \frac {r'_3}{L} \right)
^{\lambda_1/\lambda_3}
\end{equation}
 The conditions of the Theorems on page 4 require $t$ to be large enough, so
the equation (\ref{zdasymptote}) is valid for some $t>t_0$. Then
(\ref{answer}) is valid for all $t>t^*(r'_3) = -\frac 1{\lambda_3}
\ln \left( r'_3/L \right) + t_0$, or $r'_3 >L e^{\lambda_3
(t_0-t)}$. At smaller $r'_3$, the influence of the boundary has not
yet reached the region, and $\omega$ is determined by the initial
condition. So, (\ref{answer}) does not mean a real finite-time
singularity: it is naturally 'smoothed' near the center, the radius
of the smoothing part steadily decreasing. On the other hand,
(\ref{answer}) gives a power law for velocity structure functions.
Scaling (not yet multiscaling) properties are derived from the
stochastic Euler equation.

We stress that (\ref{answer}) is a long-time local approximation to any
general solution (\ref{reshenie}) of non-viscous Eq. (\ref{maineq})
with symmetric $A_{ij}$ in the regions of very high vorticity. These local
maximums of vorticity (exponentially growing, as follows from (\ref{solut})), can be interpreted
as vortex filaments. In
the next Section we discuss a simplified model that helps to
understand the solution better and to reveal the role of viscosity.
Then we proceed to the discussion on multifractality.

\section{Simple model: Details of solution and account of viscosity }

In the previous section, the long-time solution of (\ref{maineq})
was found to be nearly one-dimensional in the rotating frame
(\ref{rotation}), and its behavior appeared to be rather
deterministic. The randomness is contained mostly in the rotation
matrix $s$. Here we propose a simple one-dimensional deterministic
model equation that has similar solutions and helps to understand
the details of their behavior. Later on, we use this model to
generalize the results for finite viscosity and for multifractal
description.

The idea is to 'straighten' the random flow, excluding the matrix
$s$ and thus avoiding the need of additional rotation
(\ref{rotation}) of the frame. So we fix the random matrix $A_{ij}$
and restrict ourselves by small-scale velocity field depending on
only one variable.

Thus, consider the velocity field: $^{6}$ \footnotetext[6]{More general consideration with ${\bf u}={\bf u}(x,y,z)$ gives the same results.  }
$$
v_x= a(t) x\,,\qquad v_y = b(t) y + u(x,t)\,,\qquad v_z = c(t) z
$$
The parameters $a,b,c$ correspond to the large-scale matrix
$A_{ij}$. From incompressibility it follows
$$
a + b + c = 0
$$
The Euler equation then takes the form
$$
\begin{array}{l}
x \left( \dot{a} + a^2 \right) = - \frac{\d p}{\d x}
\\
\frac{\partial u(x,t)}{\partial t} + a(t) x \frac{\partial
u(x,t)}{\partial x} + b(t) u(x,t) + y\left( \dot{b} + b^2  \right) =
- \frac{\d p}{\d y}
\\
z \left( \dot{c} + c^2\right) = - \frac{\d p}{\d z}
\end{array}
$$
The pressure derivatives must depend linearly on $x,y,z$,
respectively. Hence, pressure must take the form:
$$
p({\bf r},t) = \frac{p_1}{2} x^2 + \frac{p_2}{2} y^2 + \frac{p_3}{2}
z^2
$$
The values $p_1$,$p_2$,$p_3$ are determined by the evolution of
$a,b,c$, respectively;  the equations for them are equivalent to
(\ref{LargeNew}). The part of the second equation that does not
depend on $y$ is:
\begin{equation} \label{onedimequation}
\frac{\partial u(t,x)}{\partial t} + a(t) x \frac{\partial
u(t,x)}{\partial x} + b(t) u(t,x) = 0
\end{equation}
This equation describes the evolution of the small-scale component
and is analogous to (\ref{maineq}).

In the region of interest velocities are small, while vorticities
are very high. So, in what follows we will discuss vorticity instead
of velocity (although very similar relations can be written for
velocity). Since $\omega(t,x) = \frac{\d u}{\d x}$, the
corresponding equation is
\begin{equation} \label{onedimomegaeq}
\frac{\partial \omega}{\partial t} + a(t) x \frac{\partial
\omega}{\partial x} -c(t) \omega = 0
\end{equation}
 We will hereafter analyze the solutions to this equation in the range
$x \in [0,1]$ for $t \ge 0$. For simplicity, let $a,c$ be constants
(although the solution can be written for arbitrary functions $a(t),
c(t)$).  Let also
\begin{equation}  \label{choicesign}
a<0 \ , \qquad b>0 \ , \qquad c=-(a+b)>b
%a<0<b<c
\end{equation}
(See Discussion for comments on the choice.) Then the values $a,b,c$
coincide with the values $-\lambda_3$, $-\lambda_2$, $-\lambda_1$,
respectively. In addition, we set the boundary condition
\begin{equation} \label{granitsa}
\omega(t,1)=1
\end{equation}
(This is a strengthened variant of (\ref{granomega}).) To provide
this boundary condition, the initial condition
$\omega(0,x)=\omega_0(x)$ must satisfy
$$
\omega_0(1)=1 \,, \quad a \frac{\d \omega_0}{\d x}(1) -c =0
$$

It is easy to check that all solutions of (\ref{onedimomegaeq}) obey
the relation:
\begin{equation} \label{sdvigu}
\omega(t,x) = e^{c(t-t')} \omega\left(t', x e^{-a(t-t')} \right)
\end{equation}
For any $x>e^{at}$,  choosing  $t'(x,t): x=e^{a(t-t')}$ we get
\begin{equation}  \label{onedimsolution}
\omega(t,x) = e^{c(t-t')} \omega\left( t',1 \right) = x^{c/a} \ ,
\quad x>\bar{x}(t) = e^{at}
\end{equation}
The value $\omega$ in this region is therefore determined by the
boundary; it is a power-law function of $x$ and does not depend on
time.

For smaller  $x$, the choice $t'=0$ gives
\begin{equation} \label{onedimcentre}
\omega(t,x) = e^{ct} \omega_0 \left( x e^{-at} \right) \ , \qquad
x<\bar{x}(t)
\end{equation}
The influence of the boundary has not spread to this inner region
yet, and the profile of $\omega$ is still determined by initial
conditions. So, there is no singularity after finite time, despite
the presence of a power law (Fig.1).
 As time passes, the size $\bar{x}$ of the inner region decreases,
 and the vorticity profile approaches the singularity but never
 reaches it.

\begin{figure}
%\vspace*{-0.37cm}
\hspace*{-0.5cm}
\includegraphics[width=9cm]{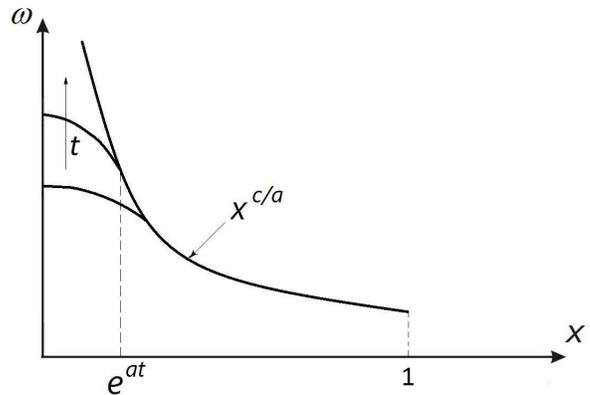}
%%\vspace*{-0.9cm}
\caption{Illustration to Eq.-s
(\ref{onedimsolution}),(\ref{onedimcentre}): dependence of
vorticity distribution on time.}
\end{figure}

 What happens if there is another boundary condition? Substituting
 arbitrary boundary condition $\omega(t,1)=f(t)$ and $t'(t,x):
 x=e^{a(t-t')}$, we have
 $$
 \omega(t,x)= x^{c/a} f \left( t - \frac 1a \ln x \right)
 \rightarrow_{t\to \infty} x^{c/a} f(t)
 $$
for any given $x$. Thus, any reasonable (i.e., slower than
exponential) function $f$ does not change the power law and
affects only the coefficient, which becomes time-dependent (see
Fig.2).
 Since the
boundary conditions correspond to large scales, the characteristic
time for $f(t)$ is of the order of the largest eddy turnover time.
This statement is also valid for the  general case (\ref{answer}).

\begin{figure}
\vspace*{-0.37cm} \hspace*{-1cm}
\includegraphics[width=10cm]{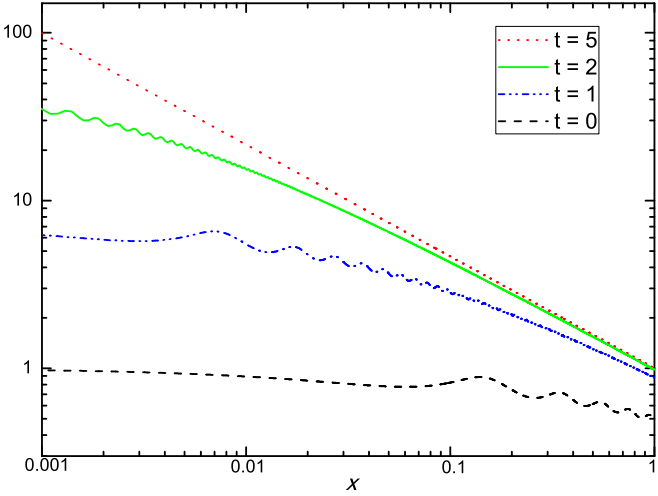} %10
%\vspace*{-0.9cm}
\caption{Evolution of vorticity distribution in one particular
case: $a=-3, b=1$, $\omega_0(x)=1/\left( 1+ \left[ x+0.1 \sin (10
\pi x) \right]^{2/3} \right)$, $\omega(t,1)=1/  \left( e^{-2t}+
\left[ 1+0.1 e^{-3t} \sin (10 \pi e^{3t}) \right]^{2/3} \right) $.
 One can see that the range of strong oscillations drifts to
smaller $x$, while inside the 'inertial range' the fluctuations
become negligible and the power law dominates.}
\end{figure}

\subsection{Evolution of spectrum}

The solution (\ref{onedimsolution}),(\ref{onedimcentre}) is not
stationary: there is always a narrowing non-stationary region
$x<\bar{x}(t)$. We now consider this solution in terms of Fourier
transform. This is useful to understand the spectrum evolution
(which gives the basis to the idea of cascade), and to take account
of viscosity.
 The Fourier transform of vorticity is:%
 $^7$\footnotetext[7]{In fact, (\ref{onedimomegaeq}) is defined for
$x\le1$. But we are interested in $k\gg 1$, so there is no
difference between Fourier series and Fourier transform. }
$$
\omega(t,k)= e^{ct} \int \limits_0^{\bar{x}(t)} e^{ikx} \omega_0
(xe^{-at}) dx + \int \limits_{\bar{x}(t)} ^{\infty}  e^{ikx} x^{c/a}
dx
$$
The first integral can be rewritten as
$$
e^{ct} e^{at} \int_0^1 e^{ikye^{at}}\omega_0(y) dy \simeq e^{-bt}
\omega_0(ke^{at})
$$
It  depends weakly
 on $k$ for all $k<\bar{k}=e^{-at}$, and
decreases exponentially as a function of time.  The second integral
is
$$k^{b/a} \int \limits_{k\bar{x}(t)} ^{\infty} e^{iy} y^{c/a} dy
$$
 It is a power law for $k<e^{-at}$ and decreases sharply % rapidly
 at larger $k$.

So, $\omega(t,k)$ is a step function of $k$, with the step running
exponentially to the right as time goes.

 To illustrate
this, we consider one particular solution of (\ref{onedimomegaeq}),
for which the Fourier transform is easy to count analytically. This
time we do not demand the 'strong' boundary condition
(\ref{granitsa}); as we have found in previous subsection, it is
enough for $\omega(t,1)$ not to grow exponentially. % ...that ... does not grow...
Let initial distribution of vorticity be
$$
\omega_0(x) = \left(1 + i x\right)^{c/a} + \left(1 - i
x\right)^{c/a}
$$
In accordance with (\ref{onedimomegaeq}),(\ref{sdvigu}) evolution of
$\omega(t,x)$ takes the form:
$$
\omega(x,t) =e^{ct}\left[ \left(1 + i e^{-at}x\right)^{c/a} +
\left(1 - i e^{-at}x\right)^{c/a}\right]
$$
$$
= 2 e^{ct}\left(1+x^2e^{-2at}\right)^{c/2a} cos\,(\phi c/a)
$$
where $\tan\,\phi = x e^{-at}$. % It coincides with (\ref{onedimomega})
For  $x \gg e^{at}$,
 we have $\phi \simeq \pi/2$, $\omega \propto x^{c/a}$.

The Fourier transform of this function is:
\begin{equation}\label{kinf}
\omega(k,t) =  |k|^{b/a}e^{-|k|e^{at}}
\end{equation}
The spectrum falls exponentially at $k\sim\bar{x}^{-1}=e^{-at}$. The
result is similar to the effect of viscosity, but the cutoff moves
along the $k$ axis towards larger values of $k$ (in case of
dissipation the cutoff would not depend on time).

Such a step spectrum spreading to larger $k$ is usually interpreted
as a cascade, or breaking of vortices. We see that in our approach
it appears without a cascade, and energy is transported to smaller
scales by means of the narrowing transition region near one selected
point, which is to become a singular point at infinite time.

It is usually assumed that viscosity is necessary to get stationary
statistical picture in turbulence. Indeed, one needs viscosity to
make all statistical averages, e.g., structure functions of all
orders, stationary: energy injected into a flow at large scales has
to be dissipated at viscous scale.

However, our example shows that, in some cases, stationary spatial
 probabilistic  distribution can be reached  even without
dissipation in some finite range of scales.

\subsection{Effect of viscosity}

It is easy to generalize (\ref{onedimequation}) to include
viscosity. Since $\nabla u$ is directed along $x$ axis,
% the viscous term is also essential in the same direction, and
the equation takes the form:
$$
\frac{\partial u(x,t)}{\partial t} + a x \frac{\partial
u(x,t)}{\partial x} + b u(x,t) = \nu\frac{\partial^2 u}{\partial
x^2}
$$
Similarly, the viscous term should  be added into the right-hand
side of (\ref{onedimomegaeq}). Changing to the new variable
$q=xe^{-at}$, we get
$$
\frac{\partial \omega(q,t)}{\partial t}  -c \omega(q,t) = \nu
e^{-2at} \frac{\partial^2 \omega}{\partial q^2}
$$
(We recall that $a<0$.) The Fourier transformation gives:
$$
\omega(k,t) =  e^{-bt} \omega_0 (k e^{at}) e^
{\frac{\nu}{2a}k^2(1-e^{2at})}
$$
It appears that, while non-stationarity produces a 'step'
(exponential fall) running to the right with exponential speed $k
\sim e^{-at}$, viscosity produces a similar (but sharper) step which
runs to the left, and very quickly (after $t\sim 1/2|a|$) becomes
stationary at $k \sim \sqrt{2|a|/\nu}$.

For the particular example of  initial condition considered in the
previous subsection,  instead of (\ref{kinf}) we then get
$$
\omega(k,t) =  |k|^{b/a}e^{-|k|e^{at}}
e^{\frac{\nu}{2a}k^2(1-e^{2at})}
$$
We see that the 'non-viscous' solution  does not differ from the
'viscous' solution in the range $k<e^{-at}$, $k<\sqrt{2|a|/\nu}$.

Although the non-viscous equation does not 'smoothen' the initial
perturbations, it  transports them to smaller and smaller scales and
multiplies by a decreasing term. So, the solutions of Euler and
Navier-Stokes equations behave similarly in the limit $t\to \infty$,
$\nu \to 0$. In this sense, the Euler equation can be treated as the
inviscid NSE.

\section{Introduction of stochastics}

In the previous section, the large-scale velocity fluctuations are
treated as deterministic ones. We have seen that this causes a
power-law dependence of vorticity. This would provide a  scaling
dependence of velocity structure functions, but it would be
mono-fractal, instead of multifractal: the scaling exponents would
be proportional to their numbers. The multifractal picture can be
restored if we take stochastic behavior of large-scale fluctuations
into account.

 The Theorems cited in page 4 claim that the 'systematic' part
 (\ref{zdasymptote}) of randomly changing matrix (\ref{qsequence})
 not only has exponentially growing averages, but also fluctuations
 of  these exponents are Gaussian random processes. An accurate
 analysis of the stochastic equation (\ref{maineq}) with account of
 these fluctuations could probably allow to construct a complete
 theory. In this paper, however, we restrict ourselves by the
 simplified example from the previous Section.

 According to the Theorems, the stochastic generalization of (\ref{onedimomegaeq})
 has the  form:
\begin{equation} \label{xiequation}
 \frac{\partial \omega}{\partial t} + (a +\xi_1(t)) x
\frac{\partial \omega}{\partial x} - (c +\xi_2(t)) \omega = 0
\end{equation}
 Here $\xi_1(t)$ and $\xi_2(t)$ are Gaussian
 random processes with zero averages.

All the relations of the previous Section can be rewritten for this
case; e.g., (\ref{sdvigu}) becomes
$$
\omega (t,x) = e^{c(t-t') +\int \limits_{t'}^t \xi_2(t'') dt''}
\omega \left( t',x e^{-a(t-t') -\int \limits_{t'}^t \xi_1(t'') dt''}
\right)
$$
For $x=0$, taking $t'=0$, we get
$$
\omega (t,0) = e^{ct +\int \limits_{0}^t \xi_2(t'') dt''} \omega
(0,0)
$$
Let $\xi_1, \xi_2$    be delta-correlated with dispersions $D_1$ and
$D_2$. (For more general case of not delta-correlated processes see
Appendix 1.) The probability density can then be written as
$$
dP[\xi_1(t), \xi_2(t)]= e^{-\frac {\int\xi_1(t')^2 dt'}{2 D_1}} e^{
- \frac{\int\xi_2(t')^2 dt'}{2 D_2}}  \prod \limits_t d\xi_1(t)
d\xi_2(t)
$$
Thus,
\begin{eqnarray}  \nonumber
\left<\omega(t,0)^n\right> &=& e^{nct} \langle e^{n\int\limits_0^t
\xi_2 dt} \rangle \omega^n(0,0)  \\
&=& e^{nct} \int e^{\int\limits_0^t \left( -\frac{\xi_2^2}{2D_2} + n
\xi_2 \right) dt} d\xi_2 (t) \omega^n(0,0) \nonumber
\\
&=& e^{nct + n^2 D_2 t/2}\omega^n(0,0)  \label{zeroav}
\end{eqnarray}
We see that the averages diverge exponentially as a function of
time.   This  characterizes the solution inside the non-stationary
inner region (\ref{onedimcentre}) with growing vorticity. The width
$\bar{x}$ of the non-stationary region is determined by the
condition
$$
\bar{x} e^{-at-\int \xi_1 dt} \simeq 1
$$
But, since $\int \xi_1 dt \propto \sqrt{t}$ after long time, we have
$at \gg \int \xi_1 dt $ and $\bar{x} \simeq e^{at}$.

Thus, adding stochastic fluctuations to $a$ and $c$ increases the
central growth  of vorticity but does not change significantly the
size of the non-stationary region.

\begin{figure*}
\vspace*{-0.37cm} \hspace*{-0.5cm}
\includegraphics[width=18cm]{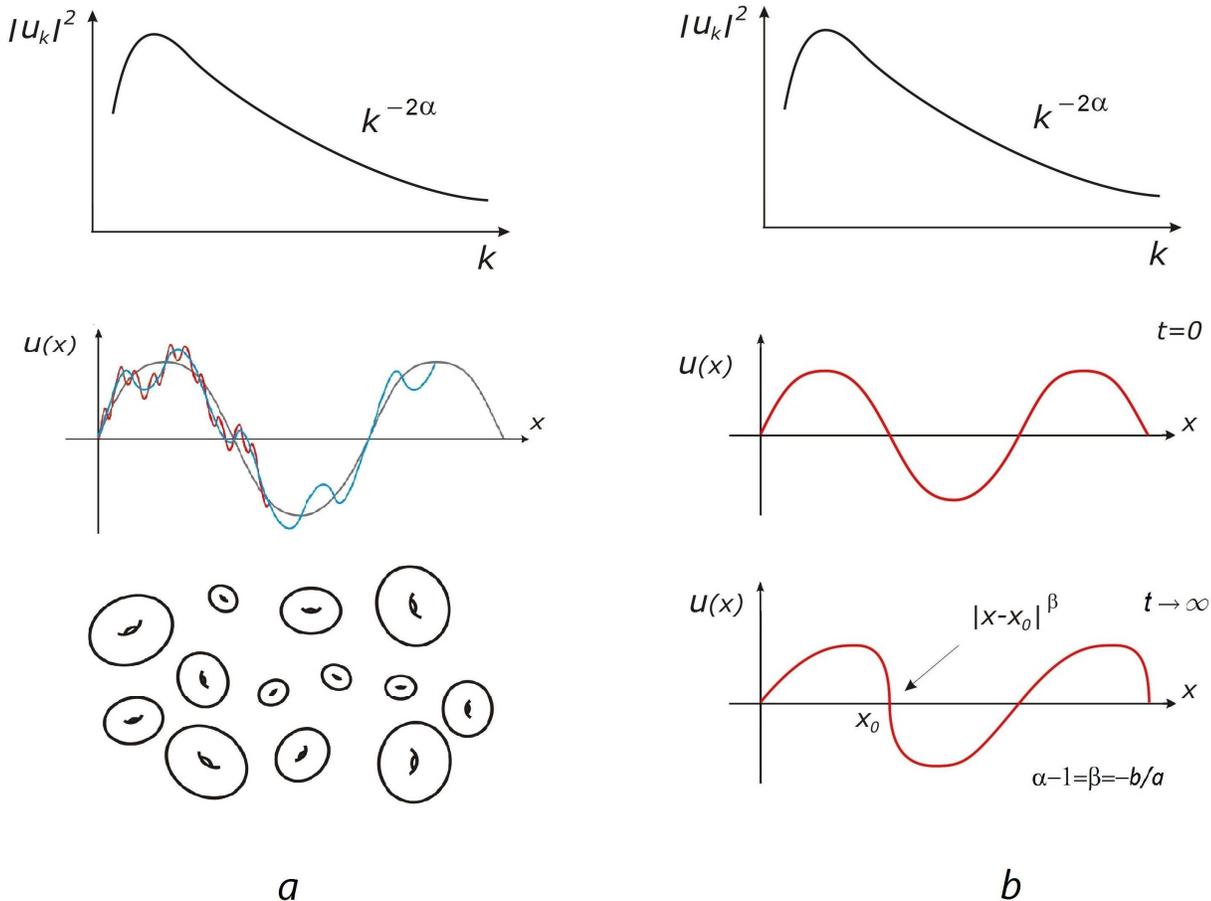}
%\vspace*{-0.9cm}
\caption{Cascade model {\it (a)} vs Infinite-time singularity ({\it
b}): the same spectrum is produced by different physical processes}
\end{figure*}

To understand the statistical properties, however, we are interested
in the  outer region $x>\bar{x}(t)$. In this case, by analogy with
(\ref{onedimsolution}), we choose $t'(x,t)$ in such a way that
\begin{equation}
\label{randboundary} \ln x = a(t-t') + \int \limits _{t'}^t \xi_1
dt''
\end{equation}
Then
\begin{equation} \label{randomom}
\omega(t,x) = e^{c(t-t') + \int \limits _{t'}^t \xi_2 dt'' }
\omega(t',1)
\end{equation}
The value $t'(x,t)$ is now a random process. An accurate
calculation of averages of (\ref{randomom}) is very complicated,
so we just make an estimate. We restrict ourselves by small $x$,
so $\left| \ln x \right| \gg 1$. An average of $\int \xi_1 dt''$
is zero, so we can estimate it by $\int \xi_1 dt'' \propto
\sqrt{t-t'} $. Thus, in (\ref{randboundary}) this term is much
smaller than $a(t-t')$ and can be neglected. From (\ref{randomom})
we then have
$$
\omega(t,x) \simeq x^{c/a} e^{\int \limits_{t-t'}^{t} \xi_2 dt''}
\omega(t',1)
$$
Raising to the power $n$ and taking an average, we get
\begin{eqnarray}  \nonumber
\langle \omega^n \rangle &= & x^{nc/a} \int e^{\int \left(
-\frac{\xi_2^2}{2D_2} + n \xi_2 \right) dt''} \prod \limits_t
d\xi_2
(t) \omega^n(t',1) \\
&\propto & x^{n \frac ca }  e^{ n^2 \frac{D_2}{2} (t-t')} \propto
x^{n \frac ca + n^2 \frac{D_2}{2a}}   \label{itogvort}
\end{eqnarray}
This scaling of vorticity moments is equivalent to velocity
structure functions with nonlinear scaling exponents:
\begin{equation}  \label{itog}
\langle \Delta v ^n (l)  \rangle \sim
 \langle \omega ^n  \rangle l^n \sim
l^{\zeta_n} \ , \quad \zeta_n
= -\frac ba n + \frac{D_2}{2a} n^2
\end{equation}
 The obtained relations provide an explanation of the  nonlinear dependence
 of  scaling exponents on their order.

The exponential average growth of vorticity in the vicinity of local maxima and power-law
distribution near these points (\ref{onedimsolution}), (\ref{onedimcentre})
produce the linear term in (\ref{itog}), while the fluctuations (\ref{randomom})
are responsible for the non-linearity.

\section{Discussion}

In this paper we derive existence and properties of vortex filaments
(high-vorticity regions) on the basis of the stochastic
Eq.(\ref{maineq}) (which in turn is derived from the NSE) by using
the Theorems (page 4). The main result is %obtaining
the scaling behavior of
vorticity (\ref{answer}),(\ref{onedimsolution}) inside these vortex
filaments, without suggesting finite-time singularities, and %deriving
multifractal behavior of vorticity (\ref{itogvort}) and velocity
({\ref{itog}) statistical moments. Thus, in our approach the direct,
not only probabilistic  formulation of the Multifractal model is
valid, however there are no singularities in the flow: at any finite
time peaks of vorticity are smoothed inside constantly narrowing
non-stationary regions.

%One  principal point is separating smaller from larger scales, which
%is not needed in the formal introduction of probabilistic
%description but is used in the 'physical' approach (Section 2).
%  The possibility of such separation is
%proved by the mechanism that provides the scaling.
In the 'canonical' cascade interpretation of the power-law spectrum,
vortices of all scales are presented and contribute to the resulting
scaling (Fig. 3a).
% In this approach it is difficult to separate the scales.
To the opposite, our consideration shows that the same
spectrum is produced by small regions near some 'almost singular'
points (Fig. 3b). Excluding these regions would cut the spectrum up
to $k\sim l^{-1}$, and the scales are easy to separate. An evidence
for this second approach comes from the numerical simulations
\cite{Farge}: the observed 'coherent structures' with high
vorticity, i.e. vortex filaments, are found to be very stable, the
lifetime exceeding many times the largest-eddy turnover time. This
contradicts to the idea of cascade. In \cite{Farge}  it is shown
that these small-scale structures are responsible for the 5/3 law.
Picking  them out breaks the power-law energy spectrum in the whole
inertial range.

The assumptions and simplifications used throughout the work are
rather general and do not seem crucial. We restricted our
consideration by symmetric large-scale velocity gradients $A_{ij}$,
supposing that the large-scale vorticity can be neglected inside the
vortex filament: the solution can probably be generalized for all
$A_{ij}$.

 Even the simplified 'straightened'
 model, apart from its illustrative functions, can be valid in the
 high-vorticity regions: the rotation matrix $s$ in (\ref{Iwasawa}), being a
 large-scale value, has a characteristic time of changes $\tau_{cor}
 \gg \omega^{-1}$, and hence its rotation can be treated as an
 adiabatic process.

Unfortunately, the numerical values of the
coefficients $\lambda_i$ (and hence the coefficients in (\ref{itogvort}), (\ref{itog}))
 are not defined by the Theorems. They depend on the
properties of the random large-scale fluctuations, and a special
investigation is needed to derive them.  However, one can consider some restrictions and
particular cases. In \cite{Falkovich}, for a similar case of polar decomposition (for which
similar theorems are valid),  the coefficient analogous to $\lambda_2$ is shown to be zero if
the probability distribution $A_{ij}$ is Gaussian. The same is true for the Iwasawa
decomposition
considered in our paper; moreover, $\lambda_2$ is zero for all large-scale
configurations that are statistically isotropic:
$$
P(A)=P(RA R^{-1})
$$
for any rotation $R$, and satisfy the  condition
% and time reversal invariant, which means
\begin{equation} \label{timerev}
P[A_{ij}(t)] = P[-A_{ij}(t)]
\end{equation}
(see Appendix 2). This requirement is stronger than the single condition of
isotropy: for example, let  matrix $A_{ij}$  take the values
$A=R\cdot \mbox{diag}(\alpha,\beta,\gamma) \cdot R^{-1}$, where $\alpha,\beta,\gamma$
are some definite (not random) quantities and $R$ is a random orthogonal matrix, $R\in SO(3)$.
  The process is isotropic if $P(A)$ does not depend on $R$.
 However, $P(A) \ne P(-A)$, since the value $-A$ is impossible.

More generally, if the distribution of traceless symmetric random matrix $A_{ij}$ is
isotropic, its probability density may depend
on only two parameters, e.g., $P(A)=P(\tr A^2, \tr A^3)$.
The additional condition $P(A) = P(-A)$ means that $P$ is an even function of its
the second argument.

The value $A_{ij}$ is defined in (\ref{seriesV}) as $\d U_i / \d
r_j$ where $\bf U$ is a velocity. Hence, the transformation $A \to
-A$ is time reversal, and the condition (\ref{timerev}) is time
conjugation invariance.

Thus, T-invariance leads to $\lambda_2=0$;  it also implies even
dependence of probability density on $\tr A^3 $, in particular,
$\langle \tr A^3 \rangle =0$. On the other hand, the large-scale
process that produces turbulence must provide some flux of energy
from outside into the flow. This breaks the T-symmetry: indeed, the
contribution of the large-scale velocity component (\ref{seriesV})
to the average energy flux through any sphere of intermediate radius
$\bar{r}\ll L$ is
%\begin{equation}
$$
\langle \Phi \rangle = \langle \int U^2 {\bf U} d{\bf s} \rangle =
\langle A_{ij} A_{im} A_{kp} \int \bar{r}_j \bar{r}_m  \bar{r}_k
\frac{\bar{r}_p}{\bar{r}} \bar{r}^2 d\Omega  \rangle
$$
$$
= \langle 2 \tr A^3 4 \pi \frac{\bar{r}^5}{3\cdot 3} \rangle   \propto
\langle \tr A^3 \rangle
$$
The condition $\langle \Phi \rangle<0$ then gives  $\langle \tr A^3
\rangle <0$, (for the 'straightened' model this means
$b=-\lambda_2>0$, compare to the signs of $a,b,c$ in
(\ref{choicesign})).

So, symmetry $t\to -t$ is forbidden if we require the income of
energy into the flow. Thus, $\lambda_2$ must not be zero, and $P(A)
\ne P(-A)$. This means, in particular,  that Gaussian probability
density is not valid for $P(A)$. (This does not contradict to the
experiments that show Gaussian behavior of large-scale velocity,
since (\ref{seriesV}) is valid for scales $l \ll L$ only.)

 However, we recall that,
%the gaussianity of large-scale fluctuations is not  required in the Theorems, while
independently on the statistics of the large-scale fluctuations,
the Theorems state that fluctuations of the exponents (\ref{zdasymptote}) are Gaussian.

Returning to the higher-order structure functions, the relation (\ref{itog})
proves their  power-law dependence and
provides an  explanation of the  nonlinear dependence
 of  scaling exponents on their order.
As it was shown in \cite{PRE2, notPRE}, the quadratic nonlinearity
describes very well  velocity scaling exponents observed in
experiments and numerical simulations. More accurate analysis of the
stochastic equations, with  account of rare events which are of most
importance for high-order structure functions, would of course add
higher degrees to the expression. But even this simplified
consideration appears to be enough to  show that average large-scale
exponents $\lambda_i$ in (\ref{zdasymptote}) determine the scaling
('fractal')  behavior of the solutions, while  fluctuations of these
exponents produce 'multifractality'.

\section{Conclusion}

Thus, in the paper we study the solutions of the Navier-Stokes
equation in the regions of high vorticity (vortex filaments),
treating the large-scale velocity fluctuations as independent
stationary random process. The stochastic equations
(\ref{maineqwithU}), (\ref{maineq})  are thus the main equations
of the paper.

We analyze the long-time asymptote of the solutions to Eq.
(\ref{maineq}) and show that an infinite-time singularity appears in
the limit $\nu \to 0, l \to 0$; for any finite $t$, there is no
singularity, and for any finite $l$ inside the inertial range, the
solution becomes a power law (\ref{answer}) after some time $t(l)$.

We show that the solution  corresponds to random rotation and
systematic exponential stretching of a vortex filament. This
exponential stretching causes the power-law distribution of
vorticity, the resulting spectrum is quite similar to that
expected from the model of breaking vortices.

Taking into account the stochastic component of the stretching, we
derive the multi-scaling distribution of vorticity (and velocity
differences), and quadratic dependence of velocity scaling
exponents on their order (\ref{itog}). As it was shown in
\cite{PRE2}, this result agrees very well with experimental and
DNS data. All these results do not depend on the assumptions on the
properties of the large-scale random process.

We are grateful to Prof. A.V. Gurevich for his permanent interest to
our work. We thank Prof. V.S. L'vov  and A.S. Ilyin for valuable
discussion and  the first anonymous referee for useful comments and
questions.

 The work was partially supported by RAS Program 18 $\Pi$.

\section{ Appendix 1: Calculation of statistical moments in the case of finite time-correlated
random process}

The averages (\ref{zeroav}), (\ref{itogvort}) are calculated under
the assumption of delta-correlated  coefficients $\xi_1, \xi_2$ in
Eq. (\ref{xiequation}). However, the same result can be obtained
 in the limit $t\to \infty$ without the assumption.
  To illustrate  this, we consider the average
  $\langle e^{k \int \limits _0 ^t a(t_1)dt_1} \rangle$
where $a(t)$ is a random Gaussian process with correlation function
\begin{equation} \label{finitecorrelator}
\langle a(t_1) a(t_2) \rangle = G(t_1-t_2) \ , \quad \int
\limits_{-\infty}^{\infty} d\tau G(\tau) =1
\end{equation}
The probability density of $a$ can be written in the form
\cite{SlavnovFaddeev}:
$$ P[a](t)= e^{-\frac 12 \int \limits_0^t dt_1 \int \limits_0^t dt_2
G^{-1}(t_1-t_2)  a(t_1) a(t_2)}
$$
where $G^{-1}$ is defined by
\begin{equation} \label{obratnayaG}
\int_{-\infty}^{\infty} dt' G(t_2-t') G^{-1}(t'-t_1)
=\delta(t_2-t_1)
\end{equation}
 For the statistical moments %of $\omega$
 we then get
\begin{eqnarray}  \label{functional}
%\langle \omega^k (t) \rangle =
&& \langle e^{k \int \limits _0 ^t a(t_1)dt_1} \rangle = \\
&& = \int \prod \limits_{\tau} da(\tau) e^{-\frac 12 \int
\limits_0^t dt_1 \int \limits_0^t dt_2 G^{-1}(t_1-t_2) a(t_1) a(t_2)
+ k\int \limits_0^t dt_1 a(t_1)} \nonumber
\end{eqnarray}
This is a Gaussian integral, thus, the saddle-point method gives an
exact result in the case \cite{SlavnovFaddeev}. The optimal
trajectory is defined by the condition:
$$
\left. \frac {\delta}{\delta a(\tau)} \right| _{a(\tau)=a_0(\tau)}
\left( -\frac 12 \int \limits_0^t dt_1 \int \limits_0^t dt_2
G^{-1}(t_1-t_2) a (t_1) a (t_2) \right.
$$
$$
\left. + k\int \limits_0^t dt_1 a(t_1) \right) =0
$$
Hence,
\begin{equation} \label{rho01}
\int \limits_0 ^t dt_1 G^{-1}(t'-t_1) a_0(t_1)= k\ , \quad 0<t'<t
\end{equation}
Making use of  (\ref{obratnayaG}), we get
\begin{equation}
\label{rho0saddle} a_0(t')= k \int \limits_0 ^t dt_1 G(t'-t_1)
\end{equation}
Substituting to (\ref{functional}) we get
%$$ %
\begin{eqnarray}    \nonumber   %\label{fullsaddle}
%\langle \omega^k (t) \rangle
\langle e^{k \int \limits _0 ^t a(t_1)dt_1} \rangle &\simeq & \exp
\left[ \int \limits_0^t a_0 (t_1) dt_1 \times \right.
\\ \nonumber
%$$
%$$
&\times& \left. \left( -\frac 12  \int \limits_0^t dt_2
G^{-1}(t_1-t_2) a_0 (t_2)  + k \right) \right]
% $$   %
\end{eqnarray}
and with account of  (\ref{rho01}), (\ref{rho0saddle}):
\begin{eqnarray} \label{momentsaddle}
%\langle \omega^k (t) \rangle
\langle e^{k \int \limits _0 ^t a(t_1)dt_1} \rangle &\simeq & \exp
\left[ \frac 12 k \int \limits_0^t dt_1 a_0 (t_1) \right] \\
&=& \exp \left[ \frac 12 k^2 \int \limits_0^t dt_1 \int \limits_0^t
dt_2 G (t_1-t_2)  \right] \nonumber
\end{eqnarray}
(One could as well obtain the  same result by shifting  $a(t)$ to
get the perfect square in the exponent.)

From (\ref{momentsaddle}) it follows that as $t\to\infty$ (or, more
precisely, for any $t$ larger than the correlation time) the
statistical moments grow exponentially, just as in the case of
delta-correlated process.

\section{Appendix 2: On the coefficients $\lambda_i$ in the case of isotropic
large-scale random process $A_{ij}(t)$. }

We consider a random isotropic traceless symmetric matrix
$A_{ij}(t)$. Denote $\gamma = q q^T$, where the matrix $q$ is
defined by (\ref{matricy}). The matrix $\gamma$ is thus a functional
of the random process $A_{ij}(t)$.

From (\ref{qsequence}), with account of $A=A^T$,  we have:
$$
 \gamma (-A) \gamma (A) = q (-A) q^T (-A) q(A) q^T(A) = I
$$
Hence, $\gamma^{-1} (A) =\gamma (-A)$. According to the Iwasawa decomposition,
$\gamma = z d^2 z^T$; in particular, $$(d^2)_{33}= \gamma_{33} \ ,$$
 $$(d^2)_{11}=
(\gamma_{22} \gamma_{33} -\gamma_{23}^2)^{-1} = \left( (\gamma^{-1})_{11} \right)^{-1}
= \left( \gamma(-A) _{11} \right)^{-1}
$$
(We recall that and $\mbox{det} \gamma = \mbox{det} q(A) = 1$.)

 For any analytic
function $f(B)$ and any rotation $R\in SO(3)$, $f(RBR^{-1})=R f(B)
R^{-1} $. Taking $f=\gamma$, \ $B=-A$, and $R=R_c= \left(
\begin{array}{ccc}   0 & 0 & -1 \\ 0 & 1 & 0 \\ 1 & 0 & 0
\end{array} \right) $, we get $\gamma (-A) = R_c^{-1} \gamma (-R_c A
R_c^{-1}) R_c $, so
$$
\left( \gamma (-A) \right) _{11} = \gamma (-R_c A R_c^{-1})_{33}
$$
Now, from Theorem 1 (page 4) we have
\begin{eqnarray}
\lambda_3 &=& \lim \limits_{N\to \infty} \langle \frac 1{2N} \log (d^2)_{33} \rangle
\nonumber\\
&=& \lim \limits_{N\to \infty} \frac 1{2N} \int \log  \left( \gamma(A)\right)_{33} P(A) DA(t)
\label{Appendlam3}\\
\lambda_1 &=& \lim \limits_{N\to \infty} \langle \frac 1{2N} \log (d^2)_{11} \rangle
\nonumber \\
=&-&\lim \limits_{N\to \infty} \frac 1{2N} \int \log  \left[
\gamma(-R_c A R_c^{-1} )_{33} \right]  P(A) DA(t) \nonumber
 \end{eqnarray}
We recall that the integral means
$\int \prod \limits_t \prod \limits_{i,j} dA_{ij} (t)  P[A(t)] $.
The last expression can be rewritten as
$$
\lambda_1 = - \lim \limits_{N\to \infty} \frac 1{2N}
\int \log \left[ \gamma (A')_{33} \right] P(-R_c^{-1} A' R_c) DA'(t)
$$
Isotropy of the distribution means $P(A)=P(A')$, $A'=RAR^{-1}$ for any $R\in SO(3)$. Taking $R=R_c$, we get
$$
\lambda_1 =-\lim \limits_{N\to \infty} \frac 1{2N}
\int \log \left[ \gamma (A')_{33} \right] P(- A') DA'(t)
$$
Comparison with (\ref{Appendlam3})  shows that if $P(A)= P(-A)$,
then $\lambda_1 = -\lambda_3 $ (and, since $\tr A=0$,
$\lambda_2=0$). This condition corresponds to a symmetry produced by
the transformation $A \to - A$. Recalling the definition of $A$
(\ref{seriesV}) one can see that this symmetry corresponds to the
time reverse.

If time reversal invariance does not hold, we have no general
relations for the values of $\lambda_i$.

%\vspace{2cm}

\end{document}